\documentclass[aps,pra,amssymb,amsfonts,floatfix]{revtex4}

\usepackage{graphicx,psfrag}
\usepackage{dcolumn}
\usepackage{bm}
\usepackage{mathbbol}
\begin{document}

\title{Randomized Dynamical Decoupling Techniques for 
Coherent Quantum Control}

\author{Lorenza Viola}
\email{Lorenza.Viola@Dartmouth.edu}
\author{Lea F. Santos}
\email{Lea.F.Dos.Santos@Dartmouth.edu}
\affiliation{\mbox{Department of Physics and Astronomy, 
Dartmouth College, 6127 Wilder Laboratory, Hanover, NH 03755, USA}}

\date{\today}

\begin{abstract}
The need for strategies able to accurately manipulate quantum dynamics
is ubiquitous in quantum control and quantum information processing.
We investigate two scenarios where randomized dynamical decoupling
techniques become more advantageous with respect to standard
deterministic methods in switching off unwanted dynamical evolution in
a closed quantum system: when dealing with decoupling cycles which
involve a large number of control actions and/or when seeking
long-time quantum information storage.  Highly effective {\em hybrid}
decoupling schemes, which combine deterministic and stochastic
features are discussed, as well as the benefits of sequentially
implementing a concatenated method, applied at short times, followed
by a hybrid protocol, employed at longer times. A quantum register
consisting of a chain of spin-1/2 particles interacting via the
Heisenberg interaction is used as a model for the analysis
throughout.\bigskip
\end{abstract}

\maketitle

\section{Introduction}

The constructive role of randomness in physical processes has been
demonstrated in various areas of research.  In stochastic resonance
\cite{Gammaitoni}, for instance, a weak signal can be amplified by the
assistance of an appropriate noise.  In quantum information
processing, noise can intensify the speed-up of quantum walks over
classical ones \cite{Kendon}, dissipation may offer new possibilities
to implement gate operations in quantum computing \cite{Beige}, while
static perturbations characterizing faulty gates can enhance the
stability of quantum algorithms \cite{Prosen}.  In quantum
communication, the use of random operations decreases the
communication cost of achieving remote state preparation and of
constructing efficient quantum data-hiding schemes~\cite{Hayden}.
Finally, random unitary operators have been recently suggested as
allowing efficient parameter estimation for open quantum
systems~\cite{EmersonSc}.

In the context of coherent quantum control, the advantages of
stochasticity have only recently been addressed
\cite{Viola2005Random,Kern2004,Kern2005,Santos2005,Kern2006,Santos2006}.
Within the framework of {\em dynamical decoupling methods}, in
particular, analytical bounds derived in~\cite{Viola2005Random} pointed to
situations where randomized protocols are expected to outperform their
deterministic counterparts in suppressing unwanted unitary dynamics as
well as decoherence in open quantum systems.  The idea of merging
together deterministic and randomized designs into {\em hybrid}
control schemes, where benefits from both approaches may be
simultaneously exploited, was also proposed in \cite{Viola2005Random}
in general control-theoretic terms, and independently validated in
illustrative situations in~\cite{Kern2005,Santos2005,Kern2006} (see also
\cite{Santos2006}).

Here, we focus on exploring the advantages of randomization in
establishing efficient control schemes for {\em arbitrary quantum
state stabilization}, that is, for engineering a quantum memory.
Efficiency is assessed in terms of both the number of control
operations needed to achieve a desired fidelity level and the rate at
which residual errors build up in the long run. We show that by
interpolating the most effective (deterministic) scheme known for
short-time decoupling with the best available (randomized) scheme for
long time, very high performance over the entire time axis may be
ensured.

\section{System and Control Setting}

Dynamical decoupling (DD) techniques have been extensively discussed
both in the original high-resolution nuclear magnetic resonance (NMR)
setting \cite{Haeberlen} and, more recently, in connection with robust
quantum information processing, see for instance
\cite{Viola98,Viola1,DDqubits} for representative contributions. 
Assume, for simplicity, an isolated finite dimensional
target system ${\cal S}$ described by a (possibly) time dependent Hamiltonian $H_0(t)$. 
The basic idea of DD is to modify the dynamics of ${\cal S}$  by adding to
$H_0(t)$ an appropriate time-dependent control field
$H_c(t)$.  The overall propagator under the total Hamiltonian and the
control propagator in units $\hbar=1$ are, respectively, $ U(t) =
{\cal T} \exp [- i\int _{0}^t (H_0 (u) + H_c (u)) du]$ and $U_c(t) = {\cal
T} \exp [- i\int _{0}^t H_c (u) du]$, where ${\cal T}$ indicates time
ordering. A transformation to a logical frame that removes $H_c(t)$ is
commonly performed, leading to a controlled evolution described by

\begin{equation}
\tilde{U}(t)=U_c^{\dagger}(t)U(t)={\cal T} 
\exp \left[- i\int _{0}^t {\tilde H_0}(u) du \right] \:,
\end{equation}
where ${\tilde H_0}(t)=U^{\dagger}_c(t)H_0(t) U_c(t)$ is the logical
Hamiltonian.  When $H_0(t)$ is {\em time-independent} and
the perturbation is {\em cyclic} with cycle time $T_c$,
{\em i.e.} $U_c(T_c)={\mathbb 1}$, where ${\mathbb 1}$ is the identity 
operator, physical and logical frame coincide
stroboscopically at $T_n=nT_c$, $n\in {\mathbb N}$.  Using the
formalism of average Hamiltonian theory (AHT), the evolution operator
in the logical frame may be expressed as 
$\tilde{U}(nT_c)=\exp (-i {\bar H_0} nT_c)$, where
 
\begin{equation}
\tilde{U}(T_c)={\cal T} \exp \left[- i\int _{0}^{T_c} {\tilde H_0}(u) du \right]
\equiv e^{-i {\bar H}_0 T_c}
\end{equation}
defines the average Hamiltonian ${\bar H_0}=\sum_{k=0}^{\infty}{\bar
H_0}^{(k)}$, with each term ${\bar H_0}^{(k)}$ computed from the
Magnus expansion \cite{ErnstBook,HaeberlenBook}.  A sufficient
convergence criterion for the series is given by $\kappa T_c<1$, where
$\kappa=||H_0||_2$ and $||A||_2=\max |{\rm eig} (A)|$, $\forall
A=A^{\dagger}$.

The above time average for ${\bar H_0}$ may be conveniently mapped
into a group-theoretic average.  In the framework of {\em bang-bang}
DD, in particular, control actions correspond to arbitrarily strong
and effectively instantaneous rotations successively drawn from a
(projective representation of a) group, ${\cal G}=\{ g_j \}$,
$j=0,\ldots, |{\cal G}|-1$.  The propagator at $T_c$ is written as
\begin{equation}
\tilde{U}(T_c)=U(T_c) = \prod _{k=0}^{|{\cal G}|-1} g_k^{\dagger} 
U_0(t_{k+1},t_k) g_k \:,
\end{equation}
which translates into a cyclic sequence of pulses $P_k=g_k
g_{k-1}^{\dagger}$, $k=1,\ldots, |{\cal G}|$, separated by intervals
$\Delta t=t_{k+1}-t_{k}$ of free evolutions, leading in turn to a cycle time
$T_c=|{\cal G}|\Delta t$.  The zeroth order contribution of the Magnus
expansion, which dominates in the limit $T_c\rightarrow 0$, is
therefore ${\bar H_0}^{(0)} = \bar{H}_{\cal G}={|{\cal
G}|^{-1}}\sum_{j} {g_j^{\dagger} H_0 g_j} $.  

The so-called {\em time suspension} is the DD goal we focus on here,
that is, we want to develop pulse sequences able to approximate
the evolution operator as close as possible to $\mathbb{1}$.
How well we succeed 
at preserving a given initial state $|\psi \rangle$
is reflected, for instance, 
in the proximity of the input-output fidelity $F(T)$ to its maximum
value 1, where $F_{|\psi \rangle}(T)=|\langle \psi |U(T)| \psi \rangle |^2$.

A deterministic protocol 
based on a {\em fixed} control path of a representation of ${\cal G}$
and aiming at achieving first-order decoupling,
${\bar H_0}^{(0)}=0$,
will be referred to as ``periodic deterministic
decoupling'' ({\tt PDD}) protocol \cite{remark}.  
We will assume here that the
first {\tt PDD} pulse occurs only at $t_1=\Delta t$, that is,
$g_0=\mathbb{1}$.
Its simplest stochastic version is
obtained by randomly picking elements over ${\cal G}$, such that the
control action at each $t_n=n\Delta t$ [$t_0$ included]
corresponds to $P^{(r)}=g_i
g_{j}^{\dagger}$, $i,j=0,\ldots, |{\cal G}|-1$.  
This leads to what we
call ``na\"{\i}ve random decoupling'' ({\tt NRD}) -- an intrinsically
acyclic method. 
Bounds on the worst-case pure-state expected fidelity
at time $T$, $F(T)= {\rm min}_{|\psi \rangle } F_{|\psi \rangle}(T)$, 
were established in Ref.~\cite{Viola2005Random}.  For {\tt
PDD}, in the limit $T T_c \kappa^2 \ll 1$, we have: 
$F(T) \geq 1 - {\cal O}(T^2 T_c^2 \kappa^4)$, 
while for {\tt NRD}
and $T \Delta t \kappa^2 <1$: ${\mathbb E}\{F(T)\}\geq 1 - {\cal O}(T
\Delta t \kappa^2)$, where ${\mathbb E}$ denotes ensemble expectation
over all control realizations.
We note that the bound for {\tt NRD} still holds in the case of a {\em
time dependent} Hamiltonian as far as $||H_0(t)||_2$ is uniformly
bounded in time by $\kappa>0$.  Within their regime of validity, these
bounds indicate that {\tt NRD} should outperform {\tt PDD} when
$|{\cal G}|^2 (T\Delta t \kappa^2) \gg 1$, which is the case when
large control groups and/or long interaction times are involved.

In practice, we avoid the extremization procedure required to
determine the worst-case pure-state 
expected fidelity. Instead, in order to get a
state-independent estimate of the performance of
different DD methods, we invoke gate entanglement fidelity,
$F_e(T)$ \cite{Schumacher}.  
This quantity is linearly related to the 
average input-output
fidelity over all pure states of the system
\cite{Nielsen,Fortunato}, and it may be computed
as $F_e (T)=|\mbox{Trace}(U(T))/d|^2$, where $d$ is the dimension
of the system's state space.  Our control
objective is then to get as close as possible to
${\mathbb E}\{F_e (T)\} \rightarrow 1$.  
In the Monte Carlo simulations to be presented 
below, ensemble expectation, ${\mathbb
E}\{F_e(T)\}$, is further replaced by an average over a sufficiently
large statistical sample of control realizations, leading to 
what we denote
$\langle \langle F_e \rangle \rangle $.

In order to concretely illustrate the benefits of randomization, we
concentrate on a relatively simple, yet physically relevant, example
-- that is, to completely refocus the internal evolution of a chain
consisting of $N$ strongly coupled spin-1/2 particles (qubits)
described by the Heisenberg model,
\begin{equation}
H_0=\sum_{i=1}^{N} \frac{\omega_i Z_i}{2} + J\sum_{i=1}^{N-1} 
\left( X_i X_{i+1} + Y_i Y_{i+1} + \Delta Z_i Z_{i+1} \right) \:.
\label{ham}
\end{equation}
Here $X,Y$, and $Z$ denote Pauli operators, $\omega_i$ is the
frequency of qubit $i$, $J$ is the coupling parameter, and $\Delta$
determines the anisotropy.  Only nearest-neighbor interactions are
considered, which is a fairly good approximation for couplings
exponentially decaying with the qubit distance -- as arising, for
instance, in quantum dot arrays \cite{Loss} -- or decaying cubically
-- as it is the case for dipolar interactions of NMR crystals and
liquid-crystals~\cite{HaeberlenBook,Baugh2005}, or electrons on
Helium~\cite{Dykman01}.

We consider qubits with approximately the same frequency $\omega_i
\approx \omega$. Accordingly, in order to remove the phase evolution
due to the one-body Zeeman terms, we perform a transformation to a
frame rotating with frequency $\omega $ and characterized by the
operator $U_R(t)=\exp [-i\omega t\sum_{i}^{N}Z_i/2 ]$.  We thus work
in a combined logical-rotating frame, whereby the effective
Hamiltonian becomes ${\tilde
H}_{R}(t)=U^{\dagger}_c(t)U^{\dagger}_R(t) [H_0(t) - \omega \sum_i Z_i/2 ] U_R(t)U_c(t)$.
While this approximation is accurate for a class of physical
systems (notably, homonuclear NMR samples), the restriction is not
fundamental. If the spread of the single-qubit is significant (so that
a common rotating frame does not exist), schemes capable of
additionally refocusing the Zeeman terms may be constructed without
adding to the overall complexity of the DD procedure~\cite{Stoll}.

\section{Convergence Improvement}

Following the general idea of \cite{Stoll,Viola1}, a
{\tt PDD} protocol capable of refocusing the nearest-neighbor
couplings of Eq.~(\ref{ham}) for arbitrary parameter values may be
built by recursively nesting DD sequences based on the group ${\cal
G}_i= \{{\mathbb 1}_i, Z_i, X_i, Y_i\}$ for each  {\em even}
qubit, $i=2,4, \ldots, 2m$, where $N=2m$ or $N=2m+1$, $m\in {\mathbb
N}$.  For instance, when $N=4$ or 5, a possible
DD scheme may be visualized in terms of the following matrix,
\[ M= \left(
\begin{array}{cccccccccccccccc}
%
\mathbb{1}& Z & X & Y 
& Y & X & Z & \mathbb{1}
&\mathbb{1}& Z & X & Y 
& Y & X & Z & \mathbb{1} \\

\mathbb{1}& \mathbb{1} & \mathbb{1} & \mathbb{1} 
& Z & Z & Z & Z
& Y & Y & Y & Y 
& X & X & X & X 

\end{array}
\right),
\]
where each row corresponds to an even qubit and each column,
supplemented with the identity operators associated to the odd qubits,
leads to an element of the DD group, so that ${\cal G}=\{g_j\}$,
$j=0,\ldots, |{\cal G}|-1$ and $g_j=\mathbb{1}_1 \otimes
[M_{(1,j+1)}]_2 \otimes \mathbb{1}_3 \otimes [M_{(2,j+1)}]_4
\otimes\mathbb{1}_5$.  Although this scheme 
does not scale efficiently, as the
number of $\pi$ pulses required to close a cycle grows as $4^m$, 
it allows for the study of the effects of
large control groups in DD methods
with no need to employ excessively large systems.
Numerical simulations with moderate computational resources
become then viable.
Contrary to {\tt PDD}, where only one qubit is rotated at a time,
the {\tt NRD} sequence associated to the above
${\cal G}$ involves random pulses ranging from 
the identity operator to $m$ simultaneous rotations.
The total number of random pulses leading
to $R$ simultaneous rotations is given by 
$Q_R=3^R m!/[R!(m-R)!]$,
where $\sum_{R=0}^{m} Q_R =4^m$.
In large systems, the percentage of random pulses
corresponding to a single qubit rotation, or at 
the other extreme, to $R=m$, is very small,
decreasing with the size of the system,
respectively, as $3m/4^m$ and $(3/4)^m $.
When $m\neq 3+4n$, the largest $Q_R$ is obtained for 
the integer $R$ 
in the interval $[(3m-1)/4,(3m+3)/4]$,  
while for $m = 3+4n$, both values 
$R=(3m-1)/4$ and $R=(3m+3)/4$ lead to the 
two largest subsets of random pulses.

\begin{figure}[t]
\begin{center}
\includegraphics[width=3.5in]{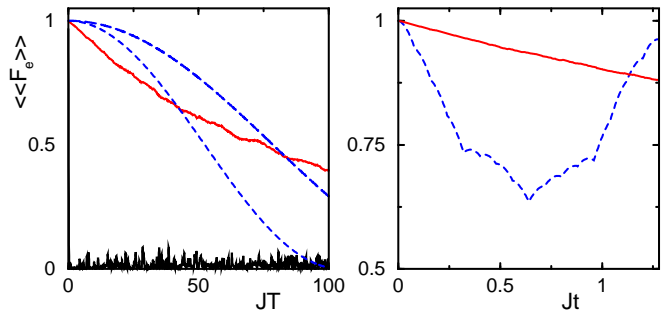}
\end{center}
\caption{(color online) {\tt PDD} vs. {\tt NRD} based on a nested
pulse sequence for Hamiltonian (\ref{ham}) with $\Delta =1$ and $N=8$
in the logical-rotating frame. Left panel: average fidelity at
$T_n=n|{\cal G}| \Delta t$, $|{\cal G}|=4^4$. {\tt PDD}:
$T_c=0.08J^{-1}$ - (blue) dashed line, $T_c=0.12J^{-1}$ - (blue)
short-dashed line; {\tt NRD}: (red) solid line, $\Delta t=
0.12J^{-1}/4^4$, average over $50$ realizations.  Free evolution:
(black) oscillating solid line.  Right panel: average fidelity within
a cycle, $t_n=n\Delta t$ with $\Delta t=0.005 J^{-1}$ for both
protocols, average over $10^2$ realizations.}
\label{fig:NN8qubits}
\end{figure}

In Fig.~\ref{fig:NN8qubits}, $N=8$ qubits are considered, leading to a
relatively large control cycle: 256 time slots.  Two situations favoring
stochastic schemes are identified. On the left panel, the average
fidelity is computed at every $T_n=n |{\cal G}| \Delta t$. Even though
{\tt PDD} achieves first-order decoupling at these instants, the
fidelity decay is substantially slower for {\tt NRD}. This behavior
persists even when the $\Delta t$ value of {\tt PDD} is shorter than
that of {\tt NRD}.  Irrespective of the validity of the strict
short-time condition underlying the bounds
of~\cite{Viola2005Random}, these findings confirm the faster
convergence offered by stochastic methods when $|{\cal G}|$ is large.
The fact that {\tt NRD} eventually surpasses both {\tt PDD} curves
shows that, for sufficiently long times, the constraints on $\Delta t$
for random DD may be relaxed or, equivalently, the number of 
control operations
able to ensure a certain fidelity level may be smaller than in {\tt
PDD}.  Both features may be very advantageous in realistic settings,
given that achievable pulsing rates are finite and excessive `kicks'
might be undesirable (leading e.g. to unwanted heating in devices
operating at dilution-refrigerator temperatures, such as quantum
dots).  Similar improvements are observed in situations where
constraints on the number of pulses or control intervals make it
unfeasible to close a complete cycle.  This may be the case, for
instance, when $T_c$ becomes prohibitively long.  Here, no analytical
fidelity bound for {\tt PDD} exists, hence we rely exclusively on
numerical simulations.  
When compared with the particular {\tt PDD} sequence considered,
{\tt NRD} performs significantly better for {\em most} intra-cycle times
$t_n=n\Delta t<T_c$, a result that prompts the search for superior deterministic
sequences. Designing new stochastic methods capable of pre-filtering
potentially good sequences simplifies this search, which would
otherwise be performed over the extremely large ensemble,
size of ${\cal O} (|{\cal G}|^{n |{\cal G}|})$, generated
by {\tt NRD}. More efficient randomized protocols will be discussed
in the next section.

\section{Long-time Improvement}

Conceptually, there are two main strategies for boosting DD
performance. One rests on ways for 
increasing the 
averaging accuracy
(hence the minimum power of $\Delta t$) in the effective
Hamiltonian, the
other on slowing down the accumulation of residual errors due to
imperfect averaging over long times.  Based on these guiding
principles, we introduce several DD schemes and discuss their relative
merits.

In the deterministic domain, one possibility is motivated by the
Carr-Purcell sequence of NMR, and consists of symmetrizing in time the
control path of the {\tt PDD}. It leads to what we call ``symmetric
deterministic DD'' ({\tt SDD}).  The cycle becomes twice as long, but
all odd order terms in ${\bar H_0}$ are also canceled. Another scheme,
which generalizes NMR supercycle techniques \cite{HaeberlenBook},
corresponds to ``concatenated DD'' ({\tt CDD}), as recently formalized
in~\cite{Khodjasteh2004}.  {\tt CDD} has a temporal recursive
structure, whose level $\ell+1$ of concatenation is determined by the
pulse sequence $C_{\ell+1}=C_{\ell} P_1 C_{\ell} P_2 \ldots C_{\ell}
P_N$, where $P_k$ is the $k$th pulse, $C_0$ is the inter-pulse
interval and $C_1$ denotes the generating {\tt PDD} sequence. At level
$\ell =2$ the concatenated sequence is also symmetric.  Interestingly,
however, {\tt CDD} may outperform {\tt SDD} even before this level of
concatenation is actually completed, reflecting its superiority in
reducing the accumulation of errors.

In terms of randomized DD, we introduce hybrid protocols,
which combine deterministic and stochastic features. The
purpose here is to ensure good performance at short times,
as typical of deterministic protocols, while at long times,
instead of accumulating errors {\em coherently}, randomization
guarantees this to happen {\em probabilistically}. The protocols
are classified
according to an {\em inner} and an {\em outer} code.
The former establishes the pulse sequence
in the interval $[n|{\cal G}|\Delta t, (n+1)|{\cal G}|\Delta t]$, 
being associated with the control path chosen to 
traverse ${\cal G}$. Since the group is 
traversed in full, as in deterministic schemes, 
the inner code leads to an effective Hamiltonian
$H_{\rm eff} \propto {\cal O}(\Delta t)$. 
The outer code
determines the additional random pulses to be applied at
$T_n=n|{\cal G}|\Delta t$. The latter are drawn from a group ${\cal G}'$,
which needs not coincide with ${\cal G}$. Randomization may
then be associated with the choice for the inner code or the outer code.
In the first category we have the ``random path decoupling''
({\tt RPD}) protocol, as proposed in \cite{Viola2005Random}.  It
consists of randomly choosing, at every $T_n$, which control path to
follow to traverse ${\cal G}$.  
Here, as in most stochastic protocols, logical and
physical frame do not always coincide and we need to keep track of
the applied control trajectory,
so that an appropriate control operation may be used to correct frames.
However, similarly to {\tt PDD}, we may choose
to fix the first group element as ${\mathbb 1}$, which leads to
frame coincidenece at every $T_n$. This may be
particularly useful, for instance, in
conventional line-narrowing spectroscopic applications. 
We will call this alternative ``pseudo-random path decoupling'' 
({\tt pRPD}). To the second
category belongs the embedded scheme ({\tt EMD}), inspired to
\cite{Kern2005}. The inner code is a fixed {\tt PDD} sequence,
while the bordering pulses may either be picked at random from ${\cal G}$
or from a different control set, corresponding, for example 
to products of uncorrelated
Pauli operators as described in \cite{Kern2005}.
Contrary to {\tt RPD}, this protocol may suffer from
non-uniform performance across the set of $|{\cal G}|!$
possible inner paths, requiring a pre-selection
of a good deterministic pulse sequence. 
The performance of both {\tt EMD} and {\tt RPD}
is significantly improved by
further symmetrizing the inner control path in the same manner
as in {\tt SDD}. Here, we will be dealing only with 
the ``symmetric random path decoupling'' ({\tt SRPD}) protocol.

\subsection{Bounds on Fidelity Decay}

Analytical upper bounds on the order of the
fidelity decay, $1-{\mathbb E}\{ F_e(T)\}$, may
give an insight on what to expect from the above protocols. Generalizing 
the arguments of~\cite{Viola2005Random,Kern2005}, we find the following
order-of-magnitude estimates for deterministic (top line) and
stochastic (bottom line) schemes:
\begin{center}
\begin{tabular}{cc}
{\tt PDD}& {\tt SDD} \\ 
$T^2 (|{\cal G}| \Delta
t)^2 \kappa^4$\hspace{0.1cm} & \hspace{0.1cm}$T^2 (|{\cal G}| \Delta
t)^4 \kappa^6$ 
\end{tabular}
\end{center}
\begin{center}
\begin{tabular}{ccc}
{\tt NRD} & {\tt RPD/EMD} & {\tt SRPD} \\ $T \Delta t
\kappa^2 $\hspace{0.1cm} & \hspace{0.1cm} $T (|{\cal G}| \Delta t)^3
\kappa^4$\hspace{0.1cm} & \hspace{0.1cm} $T (|{\cal G}| \Delta t)^5
\kappa^6$ 
\end{tabular}
\end{center}

For deterministic protocols,  residual errors add coherently,
which leads to a quadratic-in-time fidelity decay, ${\cal O}((T
||{\bar H_0}||_2 )^2)$, as found in \cite{Viola2005Random}.
Therefore, it is only the ability to cancel or reduce higher order
terms in ${\bar H_0}$ that may induce better performance.  At short
times, the dominant term in each cycle of the {\tt PDD} is ${\bar
H_0^{(1)}}$, and the bound is derived from the norm $||{\bar
H_0^{(1)}}||_2 \leq \kappa^2 T_c$.  For {\tt SDD}, ${\bar
H_0^{(1)}}=0$, and the norm of the dominant term is limited as
$||{\bar H_0^{(2)}}||_2 \leq\kappa^3 T_c^2$. In the case of {\tt CDD},
the averaging accuracy depends on the level of concatenation (level
1 recovering the results of {\tt PDD}) and on the system considered.  

Contrasted with deterministic methods, the accumulation of residual
errors for random protocols is slower, as reflected by the
linear-in-time decay of the fidelity.  This may be intuitively
justified as follows. Each step of {\tt NRD} can accumulate an error
amplitude up to $\kappa \Delta t$ and during a time $T$ there are
$T/\Delta t$ such intervals. Due to the randomization, amplitudes
add up probabilistically, leading to a decay $\propto T \Delta t
\kappa^2$.  The reasoning is similar for the other protocols, but each
step now corresponds to the total interval of the inner code, $p
|{\cal G}| \Delta t$, $p=1$ [$p=2$] for {\tt RPD/EMD} [{\tt SRPD}],
leading to ${\mathbb E}\{F_e(T)\} \geq 1 - {\cal O}( T |{\cal G}| \Delta
t ||{\bar H_{\rm eff}}||_2^2)$.  
The norm of the effective Hamiltonian is derived from the  
deterministic sequence underlying the stochastic protocol, 
the worst path being considered for {\tt RPD}/{\tt SRPD}.
Hence $||{\bar H_{\rm eff}}||_2 \sim ||{\bar H_0}||_2$
and here again $||{\bar H_0}||_2$ becomes the main
factor differentiating the protocols. We therefore expect a
significant better performance for {\tt SRPD} (whose norm comes from
{\tt SDD}) than for {\tt RPD/EMD} (whose norm is 
that of {\tt PDD}).

Merging together features of deterministic methods and pulse
randomization, as in hybrid DD schemes, suggests the possibility of
suppressing errors more effectively at both short and long interaction
times.  However, since the above bounds apply only at short times,
numerical analysis becomes necessary.

\subsection{Numerical Results}

For the model of Eq.~(\ref{ham}), first order decoupling can be achieved
through a very simple system-size-independent scheme.  It consists of
alternating two rotations around perpendicular axes, one acting on all
odd qubits, the other on the even ones, such that the cycle is closed
after 4 collective pulses.  ${\cal G}=\{ \mathbb{1},Z_1 Z_3 \ldots
Z_{N-1}, Z_1 Y_2 Z_3 Y_4 \ldots Z_{N-1} Y_{N}, Y_2 Y_4 \ldots Y_{N}\}$
is a possible DD group realization for even $N$.

\begin{figure}[tb]
\begin{center}
\includegraphics[width=3.5in]{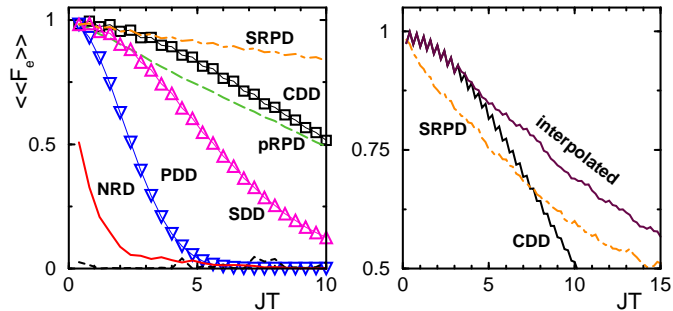}
\end{center}
\caption{(color online) Deterministic vs. random DD based on a 4-pulse
sequence for Hamiltonian~(\ref{ham}) with $N=8$ in the
logical-rotating frame.  Average fidelity over $10^2$ control
realizations, at $T_n=n|{\cal G}|\Delta t$.  Left panel: $\Delta =1$,
$\Delta t=0.1 J^{-1}$.  Free evolution: (black) dashed line at
\vspace*{-1mm}the bottom.  Right panel: $\Delta =5$, $\Delta t=0.05
J^{-1}$.}
\label{fig:NN8}
\end{figure}

A quantitative comparison is presented on the left panel of
Fig.~\ref{fig:NN8}.  Note that the (inner) sequences characterizing
{\tt SDD}, {\tt SRPD} and {\tt CDD} are not necessarily completed at
the instants of data acquisition, $T_n=4n\Delta t$. 
We verified that the 
outcomes of {\tt pRPD} and the {\tt EMD} 
based on a single group ${\cal G} $ are very similar,
while {\tt RPD} and the {\tt EMD} based on random Pauli 
operators perform closely at intermediate times.
{\tt RPD} becomes the best of the four protocols at long times.
Results are shown only for {\tt pRPD}. As
expected, random protocols surpass deterministic schemes at long
times, the crossing being evident between protocols that have
equivalent performance at short times.  We note that {\tt NRD} meets
{\tt PDD} already at very small values of 
$\langle \langle F_e \rangle \rangle$, since
the group is now small. In addition, {\tt CDD} is remarkably outperformed by
the relatively simple {\tt SRPD} method, which can be understood by
re-examining the analytical bounds. For this particular system,
{\tt CDD} at level 2 achieves only second order decoupling, so that 
${\bar H_0^{(2)}}\neq 0$ and the 
bound of {\tt SDD} is recovered. It turns out that 
due to the reducibility of ${\cal G}$, increasing 
the level of concatenation does not improve the protocol performance.
We find that ${\bar H_0^{(2)}}$ contains terms such as $X_jX_k, Y_j Y_k$,
and $Z_j Z_k$, ($j,k$ - odd),
which are unaffected by the pulses drawn from ${\cal G}$.
As a consequence, performance improvement is saturated and the 
coherent accumulation of residual errors soon deteriorates 
the results obtained with {\tt CDD} -- the method is 
eventually outperformed by {\tt SRPD}.

Given the protocols above, a way to guarantee the best performance
through the whole time axis consists of interpolating the {\tt CDD}
scheme at short times with {\tt SRPD} at long times.  
This is illustrated in the right panel of
Fig.~\ref{fig:NN8}: {\tt CDD} is used until the third level of
concatenation is reached, at $T=4^3 \Delta t$, where we then switch to
the {\tt SRPD} sequence.  Note that if the applied control history is
recorded by means of an appropriate classical register, a
randomly generated, optimized deterministic pulse sequence may be
obtained in this way upon de-randomizing the protocol at the end.

The identification of an efficient pulse 
sequence is strongly dependent on the
time interval considered. Building on the 
fact that
deterministic protocols perform better at short times, while
stochastic schemes become superior at long times, it might be
beneficial to
exploit closed-loop strategies to determine when to switch from one
to the other. While the process of monitoring a quantum system
in real time with the purpose of controlling its dynamics has 
been proposed to a variety of systems, ranging from 
cavity QED \cite{QED} to nanomechanical systems \cite{nano},
experimental implementations have been reported only recently \cite{experiments}
and the capabilities are still limited. However, 
assuming the possibility of monitoring online the system considered here, 
we could, for instance, decide when
to switch from {\tt CDD} to {\tt SRPD}  
once a certain a priori stipulated value
of $\langle \langle F_e \rangle \rangle$ is reached.

An additional advantage of randomization,
as discussed in Ref.~\cite{Santos2006},
is related to time-varying systems. In this case,
randomized protocols 
lead to more robust performance, since they
are usually more protected against adversarial situations
where a pre-established control action may be inhibited by the
system fluctuations. Here also, the possibility of real time feedback
might be useful, allowing, for instance, to 
better adjust the pulse sequence based on the system
parameters variations.

\section{Conclusions}

We have reinforced the advantages of randomization in terms of faster
convergence and long-time stabilization, by comparing the performance
of various decoupling schemes in refocusing the evolution of a chain
of nearest-neighbor-interacting qubits.  Our analysis also indicates
the promising role of stochasticity in the search of optimized pulse
sequences.  While preliminary results indicate that the main
conclusions remain unchanged when pulse imperfections are considered,
further analysis is needed in this direction.  It is also our hope
that these findings will prompt experimental verifications in
available control devices.

\section{Acknowledgements} 
\noindent

We thank Oliver Kern for a careful reading of
the manuscript and useful suggestions. 
L.V. thanks Victor S. Batista for the opportunity to attend and
contribute to a PQE stimulating session on Multipulse Coherent Quantum
Control.  Partial support from Constance and Walter Burke through
their Special Projects Fund in Quantum Information Science is
gratefully acknowledged.


\begin{thebibliography}{99}
\bibitem{remark} Note that the norm of the resulting effective Hamiltonian,
$H_{\rm eff}$, is of ${\cal O}(\Delta t)$. In general, the $k$-th order
of DD is characterized by averaging accuracy of ${\cal O}((\Delta t)^k)$.
\end{thebibliography}
\end{document}